 \documentclass[conference,final,twocolumn]{IEEEtran}
\usepackage{mathrsfs}
\usepackage{algorithm}
\usepackage{algorithmic}
\usepackage{graphicx}
\usepackage{cite}
\usepackage{citesort}
\usepackage{color}
\usepackage{psfrag}
\usepackage{subfigure}
\usepackage{amssymb}
\usepackage{epsfig}
\usepackage{pifont}
\usepackage{amsmath}
\usepackage{array}
\usepackage{multicol}
\usepackage{multirow}
\usepackage{pifont}
\usepackage{indentfirst} 
\usepackage{amsfonts}
\usepackage{fancyhdr}
\usepackage{amscd}
\usepackage{bm}
\usepackage{tikz}
\usepackage{xcolor}
\usepackage{etoolbox}
\usetikzlibrary{chains,fit,shapes}

\usetikzlibrary{decorations}
\usetikzlibrary{decorations.pathreplacing}
\usetikzlibrary{calc}
\usetikzlibrary{arrows}

\newtoggle{quickdecim}

\newtheorem{proposition}{Proposition}

\IEEEoverridecommandlockouts
\begin{document}
	
	\title{Energy-saving Pushing Based on Personal Interest and Context Information}
\author{
\IEEEauthorblockN{{Chuting Yao, Binqiang Chen and Chenyang Yang}}
\vspace{0.2cm}
\IEEEauthorblockA{Beihang University, Beijing, China\\
Email: \{ctyao,chenbq,cyyang\}@buaa.edu.cn}
\and \IEEEauthorblockN{{Gang Wang}} \vspace{0.2cm}
\IEEEauthorblockA{NEC Labs, China \\
Email: wang\_gang@nec.cn}
}

\thanks{This work is supported by China NSFC under Grant 61120106002, 973 Program  under Grant 2012CB316003, and NEC Labs.}

\maketitle

\begin{abstract}
Pushing files to users based on predicting the personal interest of each user may provide higher throughput gain than broadcasting popular files to users based on their common interests. However, the energy consumed at base station for pushing files individually to each user is also higher than broadcast. In this paper, we propose an energy-saving transmission strategy for pre-downloading the files to each user by exploiting the excess resources in the network during off-peak time. Specifically, a power allocation and scheduling algorithm is designed aimed to minimize the extra energy consumed for pushing, where network and user level context information are exploited.
Simulation results show that when the energy of both  content placement and content delivery is taken into account, the proposed unicast strategy consumes less energy and achieves higher throughput than broadcasting when the files popularity is not uniform and the personal interest prediction is with less uncertainty.

\end{abstract}

\begin{keywords}
Pushing, unicast and broadcast, energy-saving
\end{keywords}

\section{Introduction}
Since a large amount of
traffic is generated by a few popular contents, proactively caching contents at base stations (BSs) or even at users can reduce backhaul cost  \cite{NAA13} and end-to-end delay \cite{HS14}, improve energy efficiency
\cite{LY14,BWZ12,YCL14}, and boost network throughput by offloading \cite{NAA13,WZL14,CY15}, which is a promising way to support the explosively increasing traffic demands of fifth generation (5G) cellular networks with low cost.

Proactive caching consists of \emph{content placement} before the actual demand arrives and \emph{content delivery} after the user requests the files. Placing contents to users according to the predicted demands can offload wireless  traffic in peak time to off-peak time. If a user can find its requested contents in local cache, it is no need for a BS to deliver the contents to the cache hit user. Otherwise, the content delivery consumes the spectrum and the energy resources at the BS.

With the predicted personal interest of each user, pushing files to an individual user has long been regarded as a technique for improving the user experience \cite{HFG12}, where the
content placement is accomplished by \emph{unicast} when the channel of the user is in  good condition \cite{HFG12,LSZ10}. With the predicted common interest of the users in a cell (i.e., file popularity), the same set of popular contents can be pre-cached at the users by \emph{broadcast} \cite{WZL14,CY15}. 
Because the demand statistics of the users in a cell differ from that of each user \cite{HS14}, pushing the possibly interested files to a user during the off-peak time before
its actual demand arrives may provide higher throughput and lower energy in content delivery than broadcasting popular contents to all users thanks to the higher hit rate. However, because unicast has to be applied due to the different interests of multiple users, the content placement by pushing may consume more energy than broadcast. To enjoy the high throughput gain of pushing without incurring high cost, which can be captured by the energy consumed at the BS for both content placement
and content delivery, it is important to develop energy-saving  transmission strategy for pre-downloading the files to each user.

In this paper, we strive to optimize power allocation and user scheduling that minimizes
the energy consumed by the BSs for the unicast content placement. In order to exploit the excess network resource, we develop an algorithm based on the user and network level context information.  Simulation results demonstrate that the proposed strategy provides high network throughput and low energy consumption than broadcasting popular files to all users in a cell, where the energy consumed by both content placement and content delivery is taken into account, especially when the  average arrival rate of the content delivery request is high.


\vspace{-1mm}

\section{System Model}
Consider $M$ small cells hexagonally  deployed in a macro cell, where each small BS (SBS) is equipped with $N_t$ antennas.  The maximal transmit power of each SBS is $p_{\max}$, and the maximal bandwidth is $W_{\max}$. 
\subsection{Traffic Model}
Consider that the SBSs serve $K$ randomly arrived MSs requesting files of $B$ bits in total from a catalog with a time-slotted fashion, where the duration of each time slot is
$\Delta_t$. At the same time, a given portion of the resources are reserved for each real-time (RT) request such as phone call, which has high priority. The  content delivery can only use the residual resources at each SBS, which is time-varying. Denote the transmit power and bandwidth occupied by the RT traffic of the $i$th SBS in the $t$th time slot as $p^t_{i,\rm RT}$ and $W^t_{i,\rm RT}$.

The arrival of the content delivery requests is often bursty owing to the human behavior, which exhibits peaks (e.g., during the lunch  and dinner times). To capture this feature, we assume that the requests of content delivery only arrive at
the peak time. For mathematical tractability, assume that only one mobile station (MS) can be served by a SBS in each time slot and the MS only accesses to the closest SBS. Denote $s^t_{i,k}\in \{1,0\}, i=1,\ldots, M, k=1,\ldots,K$ as the scheduling indicator. When the $i$th SBS (denoted as BS$_i$) schedules the $k$th user (denoted as MS$_k$) in the $t$th time slot, $s^t_{i,k}=1$, otherwise $s^t_{i,k}=0$. Then, $\sum_{i=1}^M s^t_{i,k} \leq 1$ and $\sum_{k=1}^K s^t_{i,k} \leq 1$.

Assume that the  macro BS (MBS) can predict the interest of each MS, and gather the user and network level  context information \cite{yao2015context} from each SBS and each MS. With these information, the MBS can make the resource usage plan for each SBS to pre-download (also refer to as push in the sequel) the possibly interested contents to each MS during the off-peak time before the MS's request arrives.


\subsection{Channel Model}
We divide the off-peak time into $T_f$ frames. Each frame is further divided into $T_s$ time slots. Hence, the entire off-peak time contains $T\triangleq T_f T_s$ time slots.
Due to the user mobility, the large-scale fading gain may vary among different frames. The small-scale fading is modelled as block fading, which may vary among time slots and remains constant in each time slot.

The received signal of MS$_k$ in the $t$th time slot is
\begin{equation}\label{E:signal}
\textstyle y^t_k = \sum_{i=1}^M s^t_{i,k}\sqrt{\alpha^{\lceil\frac{t}{T_s}\rceil}_k} ({\bf h}^t_k)^H {\bf
	w}^t_k\sqrt{p^t_k} x^t_k
+ n^t_k,
\end{equation}
where $x^t_k$ is the transmit symbol with
$\mathbb{E}\{|x^t_k|^2\}=1$,
$p^t_k$ is the transmit power, ${\bf{w}}^{t}_k \in \mathbb{C}^{N_t \times 1}$ is
the beamforming vector, ${\bf{h}}^{t}_k \in \mathbb{C}^{N_t \times 1}$ is the
independent and identically distributed (i.i.d.) Rayleigh fading
channel vector, $\alpha_k^{\lceil\frac{t}{T_s}\rceil}$ is the corresponding
large-scale fading gain between the user
and the closest SBS, and $n^t_k$ is the noise with variance $\sigma^2$. $\mathbb E\{\cdot\}$ represents expectation, and ${\lceil\cdot\rceil}$ is the
ceiling function. Given that MS$_k$ is scheduled only by one SBS in each time slot, maximum ratio transmission is optimal, i.e., ${\bf{w}}^{t}_k = {\bf{h}}^{t}_k /\|{\bf{h}}^{t}_k
\|$, where $\|\cdot\|$ denotes Euclidean norm.

In the $t$th time slot, the achievable rate of MS$_k$ in nats is\vspace{-1mm}
\begin{equation}\label{E:Rate}
R^t_k =  \textstyle \sum_{i=1}^M s^t_{i,k} W^t_i \ln(1+g^t_k p^t_k),
\end{equation}
where $W^t_i \triangleq W_{\max}- W^t_{i,\rm RT}$ is the residual
bandwidth available for pushing files at BS$_i$ in the $t$th time slot,
$g^t_k \triangleq{
	\alpha_k ^{\lceil\frac{t}{T_s}\rceil}\|{\bf h}^t\|^2}/{ \sigma^2} ={
	\alpha_k ^{\lceil\frac{t}{T_s}\rceil}\|{\bf h}^t\|^2}/{( N_0 W^t_i)} $ is the
equivalent channel gain and $N_0$ is noise power spectrum density.

\vspace{-1mm}
\subsection{Power Model}
Assume that a SBS can be switched into sleep mode when the SBS has no traffic to serve. The pre-downloading power consumed by the SBSs at the $t$th time slot contains the transmit power for multiple users and the \emph{extra} circuit power consumed for operating the SBS to push the files, which can be modeled as \cite{earth2010},
\begin{align}\label{E:PowerModel_NRT}
\!\! p^t_{{\rm C}}= &\textstyle \sum_{k=1}^K \tfrac{1}{\xi} p^t_{k} +
\nonumber \\ &\textstyle  \sum_{i=1}^M  {\bf 1}(p^t_{i, \rm RT}=0){\bf 1}(\sum_{k=1}^{K} s^t_{i,k}>0)( p_{{\rm act}}-p_{{\rm sle}}),
\end{align}
where  $\xi$ is the power amplifier efficiency, $p_{\rm act}$ and
$p_{\rm sle}$ are the circuit power consumptions when the SBS is in active and
sleep modes, respectively, and ${\bf 1 }(x)=1$ when the event $x$ is true,
otherwise, ${\bf 1 }(x)=0$.

\vspace{-1mm}
\section{Energy-saving Pre-downloading}

Based on the predicted personal interests of each user, the SBSs can push the contents to each user during the off-peak time to exploit the excess spectrum resource. However, energy will be the wasted if the pre-downloaded files are not requested by the users. To circumvent this problem, we propose a context-aware transmission strategy for pre-downloading that minimizes the extra energy consumed for pushing.  For easy elaboration, we start by single-user case, and then extend the strategy to multi-user case.

\subsection{Single-user Pre-downloading}


We first optimize the transmission strategy to pre-download the $B$ bits contents to one user (say MS$_k$) in the off-peak time with duration $T$.
Specifically, we optimize $s^t_{i,k}$ and $p^t_k$ in each time slot. For notational simplicity, we omit the subscripts $i$ and $k$ in this subsection. Since the user is only accessed to its closest SBS in each time slot, $s^t=1$ when $p^t>0$ and $s^t=0$ otherwise. Hence, we only need to optimize $p^t$.

If all the information of $g^t, W^t_{\rm RT}, p^t_{\rm RT}, t=1,\cdots, T$ is available at the MBS  in off-peak time,  the MBS can optimize the resource allocation for the SBS to minimize the pre-downloading energy consumption from the following problem, \vspace{-2mm}
\begin{subequations}\label{P:1}
	\begin{align}\label{P:0}
\!\!\!	\min_{p^1,\cdots, p^T}~& \!\!\!\textstyle \frac{1}{T}\!\sum\limits_{t=1}^T\!\! \big(\tfrac{1}{\xi}p^t+
	{\bf 1}( p^t_{\rm RT}=0){\bf1}( p^t>0)(p_{\rm act}-p_{\rm sle})\big)
	\\ \label{P:1b}
	s.t. ~& \textstyle\frac{1}{T}\sum\nolimits_{t=1}^T W^t \ln (1+g^t p^t) =
	\frac{B\ln 2}{T \Delta_t},
	\\ \label{P:1a}
	&p^t\geq0,p^t+p^t_{\rm RT}\leq p_{\max},t=1,\ldots,T,
	\end{align}
\end{subequations}
where $p^1,\ldots,p^T$ is the power allocated to the user
during all the $T$ time slots. \eqref{P:1b} is the transmission rate
constraint of pre-downloading $B$ bits within $T$ time slots, \eqref{P:1a} is the power constraint of the SBSs.

The optimal solution of problem \eqref{P:1} satisfies the following multi-level water-filling  structure \cite{yao2015context}
\begin{align}\label{E:powerallocation_perfect}
\textstyle  p^{t*}  =\!\!\left\{\!\!\!\! {\begin{array}{*{20}{c}}
	{\big(\frac{W^t}{W_{\max}}\nu ^*- \frac{1}{g^t}\big)_{0}^{p_{\rm
				max}-p^t_{\rm RT}},}&{\!\!\!\!t\in {\mathcal T}_{\rm oc}\cup \mathcal N^{*} }\\
	{0,}&{\!\!\!\!t\in{\cal T}_{\rm id}-\mathcal N^{*} }
	\end{array}}, \right.
\end{align}
where $\nu^*$ is the water-filling level, $\mathcal T_{\rm oc}= \{t|p^t_{\rm RT}>0\}$ or $\mathcal T_{\rm id} = \{t|p^t_{\rm RT}=0\}$ is the index set of the time slots that have or not  have the RT traffic, and ${\cal N}^* = \{t|g^t\geq g^*_{\rm th},t\in {\cal T}_{\rm id}\}$ is index set of the scheduled
$N^*$ time slots without RT traffic for pre-downloading, which is determined by a threshold $g^*_{\rm th}$, and the function $(x)^a_0$ means  $\max\{\min\{a,x\}\}$.
Since $p^{t*}\geq 0, t \in \mathcal N^{*}$ holds for any $g^t \ge g^*_{\rm th}$, we have,
\begin{align}\label{E:QoS_nu_gth2}
\textstyle \nu^*-\frac{1}{g^*_{\rm th}}\geq 0.
\end{align}

In practice, due to the user mobility and random arrival of the RT service, the information of $g^t$, $W^t$ and $p^t_{\rm
RT}$ in all the $T$ time slots is hard to know, especially the small
scale fading gains and residual resources available for pre-downloading
at each SBS. Fortunately, the information in all time slots is only needed when computing the water-filling level $\nu^*$ and  the threshold $g^*_{\rm th}$ \cite{yao2015context}.
If the MBS can estimate $\nu^*$ and  $g^*_{\rm th}$ from the following context information and then inform the SBS  closest to the user, the SBS can allocate the power using \eqref{E:powerallocation_perfect} in the $t$th time slot only with the knowledge of $g^t$, $W^t$ and $p^t_{\rm RT}$ in this time slot.

\begin{itemize}
	\item \emph{Network level context information}:
	
	From the statistics of traffic loads in the past, the resource utilization status of a SBS can be estimated. The statue can be modelled as a probability that   $\left(1-\tfrac{l}{L}\right)\cdot 100\%$ bandwidth is occupied by RT traffic in each frame, i.e., ${\rm P}_{l}^{j} \triangleq {\rm Pr}(W^t = \tfrac{l}{L}W_{\max}),t=1+(j-1)T_s,\ldots,jT_s,j=1,\ldots,T_f$, $l\in \{0,1,\ldots,L\}$,
	where $L$ is the maximal number of RT requests that a SBS can support in one time slot. The reserved transmit power for RT service is assumed in proportion to the occupied bandwidth, i.e., $p_{\rm RT}^t =(1- \tfrac{W^t}{W_{\max}})p_{\max}$. We assume that ${\rm P}_{l}^{j},l = 0,\ldots,L$ are known at the SBS, which is reported to the MBS as the network level context information. Note that only ${\rm P}_{L}^{j} $ (i.e., the probability that a SBS is not occupied by RT traffic) is assumed known  in \cite{yao2015context}.
	\item \emph{User level context information}:
	
	From the predicted trajectory of the user, the large scale fading gains can be obtained from the radio map
 as $\alpha^1,\ldots,\alpha^{T_f}$ during the $T_f$ frames \cite{abou2014toward}. Consider $g^t = {\alpha^{\lceil\frac{t}{T_s}\rceil}\|{\bf h}^t\|^2}/{( N_0 W^t)}$ and denote $\tilde g^t \triangleq {\alpha^{\lceil\frac{t}{T_s}\rceil}\|{\bf h}^t\|^2}/{( N_0 W_{\max})}$, then the equivalent channel gain can be rewritten as $g^t =  \tfrac{W_{\max}}{W^t}\tilde g^t$.	
 For Rayleigh fading channel, $\tilde g^t$
 in the $j$th frame follows Gamma distribution, whose probability density function (pdf) is \vspace{-1mm}
	\begin{align}\label{E:distribution}
	\textstyle f^j(g) = \frac{1}{\Gamma(N_t)}\left(\frac{N_0W_{\max}}{\alpha^j}g\right)^{N_t-1}\exp\left({-\frac{N_0W_{\max}}{\alpha^j}g}\right),
	\end{align}
which is assumed known at the SBS, and is reported to the MBS as the user level context information.
\end{itemize}

To estimate the water-filling level and threshold with the two levels of context information, we transform problem  \eqref{P:1} into another problem
by using the relation of  $p^{t*}, t=1,\cdots, T$ with $\nu^*$ and  $g^*_{\rm th}$ in \eqref{E:powerallocation_perfect}, and consider the case where the small-scale fading is ergodic  in each  frame.

In what follows, we separately transform the objective function and constraints of problem \eqref{P:1}.

By substituting \eqref{E:powerallocation_perfect} into \eqref{P:0}, we can obtain the following proposition.
\begin{proposition} \label{Pro:1}
	When $\nu\leq p_{\max}$ and the small scale fading and available bandwidth are ergodic in each frame, the objective function in \eqref{P:0} becomes\vspace{-2mm}
	\begin{align}\label{E:Pro1}
	&\textstyle\frac{1}{\xi}\frac{1}{T_f}\sum_{j=1}^{T_f}\sum_{l=1}^{L-1}{\rm P}_{l}^j\frac{l}{L} \int_{\tfrac{1}{\nu}}^{\infty}\left(\nu -\frac{1}{g}\right) f^j(g) {\rm d} g+
\nonumber 	\\ &\textstyle
	\frac{1}{\xi}\frac{1}{T_f}\sum_{j=1}^{T_f}{\rm P}_L^j\int_{g_{\rm th}}^\infty \left(\nu -\frac{1}{g}\right)f^j(g) {\rm d} g+
\nonumber \\&\textstyle	\frac{p_{\rm act}-p_{\rm sle}}{T_f}\sum_{j=1}^{T_f}{\rm P}_L^j \int_{g_{\rm th}}^\infty f^j(g){\rm d}g.
	\end{align}	

\end{proposition}\vspace{-1mm}

Using similar derivations, we can obtain Proposition 2.
\begin{proposition}\label{Pro:2}
	When $\nu\leq p_{\max}$ and the small scale fading and available bandwidth are ergodic in each frame, constraint \eqref{P:1b} becomes \vspace{-1mm}
	\begin{align}\label{E:QoS_nu_gth1}
	& \textstyle\frac{1}{T_f}\sum_{j=1}^{T_f}\sum_{l=1}^{L-1}{\rm P}_l^j \tfrac{l}{L}W_{\max}\int_{\tfrac{1}{\nu}}^{\infty}\ln(\nu g)f^j(g){\rm d} g +
	\nonumber \\ &\textstyle
	\frac{1}{T_f}\sum_{j=1}^{T_f}{\rm P}_L^j W_{\max}\int_{g_{\rm th}}^\infty \ln(\nu g)f^j(g) {\rm d} g = \textstyle \frac{B\ln 2}{T\Delta_t}.
	\end{align}
\end{proposition}


The maximal and minimal power constraints in  \eqref{P:1a} are guaranteed by the function $(\cdot)_{0}^{a}$  in \eqref{E:powerallocation_perfect}.
Then, the water-filling level and threshold can be estimated from the following optimization problem, \vspace{-2mm}
\begin{subequations}\label{P-T}
	\begin{align}
{\bf P1:~}&	\min_{ \nu, g_{\rm th}} ~\eqref{E:Pro1} \nonumber
	\\ & s.t. ~	\eqref{E:QoS_nu_gth1},\eqref{E:QoS_nu_gth2}\nonumber
	\end{align}
\end{subequations}
Denote the solution of problem {\bf P1} as $ \hat \nu^*$ and $ \hat g^*_{\rm th}$.
Problem {\bf P1} is equivalent to problem \eqref{P:1} when channel and available bandwidth are ergodic and when the solution of \eqref{P:1} satisfies $\nu^*\leq p_{\max}$,\footnote{The solution of problem \eqref{P:1} satisfies this condition easily since the duration $T\Delta_t$ for pre-downloading $B$ bits is long.} in the sense that the optimal solutions of $\nu$ and $g_{\rm th}$ obtained from the two problems are identical.

In {\bf P1}, the objective function and constraints are differentiable with respect to the variables $ \nu$ and $ g_{\rm th}$ since they are the integration of continuous functions. Hence, the optimal solution must satisfy the \emph{Karush-Kuhn-Tucker}(KKT) conditions \cite{Boyd}. From the KKT conditions of problem {\bf P1}, we can prove Proposition 3. The proof is omitted due the space limitation.
\begin{proposition}\label{Pro:3}
The optimal water-filling level $\hat \nu^*$ and threshold $\hat g^*_{\rm
th}$ satisfy the following equation,\vspace{-2mm}
\begin{align}\label{E:nu_gth}
\!\!\!\!   ( \hat \nu^* - \tfrac{1}{ \hat g^*_{\rm th}}) + \xi (p_{\rm act}-p_{\rm sle})
   -\hat  \nu^*\ln(\hat \nu^* {\hat  g^*_{\rm th}}) = 0
\end{align}
and $\hat \nu^* {\hat  g^*_{\rm th}}>1$.
\end{proposition}

Further considering that the optimal solution should satisfy the constraint in \eqref{E:QoS_nu_gth1}, $\hat \nu^*$ and $\hat  g^*_{\rm th}$ satisfy two equalities in \eqref{E:QoS_nu_gth1} and \eqref{E:nu_gth}.
By taking the derivation of \eqref{E:nu_gth} with respect to $\hat  g^*_{\rm th}$ and considering $\hat \nu^* {\hat  g^*_{\rm th}}>1$ in Proposition \ref{Pro:3}, we have
$
\frac{\partial  \hat \nu^*}{\partial\hat  g^*_{\rm th}} ={\frac{1}{\hat g^*_{\rm th}}\big(\frac{1}{\hat  g^*_{\rm th}}-\hat \nu^*\big)
}/{\ln(\hat  \nu^* \hat  g^*_{\rm th})}<0
$, i.e., $\hat \nu^*$  is a monotonic decreasing function of $\hat g^*_{\rm th}$. Using the similar way, we can show that the left hand side of \eqref{E:QoS_nu_gth1} is a  monotonic decreasing function of $\hat  g^*_{\rm th}$. This suggests that we can find the global optimal solution of problem  {\bf P1} by a
two-tier bisection searching algorithm. In the inner tier, we find $\hat \nu^*$ with given $\hat g^*_{\rm th}$ by bisection searching from \eqref{E:nu_gth}. In the outer tier, we find $\hat g^*_{\rm th}$ by bisection searching from \eqref{E:QoS_nu_gth1}.

%
%
%

The single-user pre-downloading strategy during the off-peak time can be summarized as the following two steps.
\begin{enumerate}
	\item  The MBS estimates $\hat \nu^*$ and $\hat  g^*_{\rm th}$ using the two-tier bisection searching algorithm based on the context information. Since the information is in long term, the role of such an estimation is to make the resource planning.
	\item In the $t$th time slot, with the instantaneous information of $g^t, W^t_{\rm RT}, p^t_{\rm RT}$, the SBS optimizes the transmit power based on \eqref{E:powerallocation_perfect} with the estimated water-filling level and threshold, and then $s^t$ can be obtained. When $p^{t*} =0$, the user will not be scheduled by the SBS.
\end{enumerate}
\vspace{-2mm}
\subsection{Multi-user  Pre-downloading}
When targeting at minimizing the total energy consumption for pre-downloading files to multiple users, scheduling and power allocation need to be jointly designed, whose optimal solution is hard to find due to the coupling among  users. In the sequel, we propose a heuristic strategy, where the user scheduling and power allocation are separately designed.

Inspired by the single-user strategy, the power allocation to the $T$ time slots for each MS is designed by using context information. Since there may exist multiple MSs in each cell, the resource
utilization status for each MS should also reflect the resources occupied by pushing contents for other MSs. The pre-downloading strategy for multiple users can be implemented as follows.
\begin{enumerate}
	\item \emph{Power Allocation}:
	After gathering the predicted trajectories of all MSs in the macro cell, the MBS is aware of the  set of users located in the $i$th small cell in the $j$th frame (denoted as ${\cal K}_i^j$ with cardinality $K^j_i$). By assuming that BS$_i$ selects one of the users in ${\cal K}_i^j$ with same probability, the time resources (i.e., time slots) available for each user in ${\cal K}_i^j$ will reduce $K^j_i$ times. Therefore, for MS$_k$, the resource utilization status of its closest SBS (say, BS$_i$) can be modelled as,\vspace{-2mm}	
		\begin{equation}
		\textstyle \hat {\rm P}_{k,l}^j= 	\frac{{\rm P}_{k,l}^j}{K^j_i},	\quad l=0,\ldots,L, 	\quad k\in{\cal K}_i^j
		\end{equation}
with which the MBSs can estimate the water-filling level and threshold  for MS$_k$ as $\hat \nu_k^*$ and $\hat g_{{\rm th},k}^*$ from problem {\bf P1}. Then, in the $t$th time slot, which is in ${\lceil\frac{t}{T_s}\rceil}$th frame, BS$_i$ can optimize the power to the users in ${\cal K}_i^{\lceil\frac{t}{T_s}\rceil}$ according to their own water-filling level $\hat \nu_k^*$ and threshold $\hat g_{{\rm th},k}^*$, as well as their own instantaneous channel $g_k^t$ and available resources $W_{\max}-W^t_{i,\rm RT}$ and $p_{\max} -p^t_{i,\rm RT}$ by using \eqref{E:powerallocation_perfect}.
	
\item \emph{Multi-user Scheduler}: In the $t$th time slot, according to the estimated water-filling levels and thresholds, the set of MSs in the $i$th cell who are allocated with non-zero powers is defined as a conflict set (denoted as $\hat {\cal K}_i^{t}$ with cardinality $\hat K^t_i$). To ensure the fairness among the MSs in using the residual resources in each SBS, random scheduler is employed, i.e., the MSs in $\hat {\cal K}_i^{t}$ is scheduled by BS$_i$ with the same probability of $\frac{1}{\hat K^t_i}$.

\end{enumerate}
%



\vspace{-2mm}

\section{Simulation Results}\vspace{-1mm}
In this section, we evaluate the network throughput and energy consumption of the proposed pre-downloading strategy, where the energy consumed both by content
placement and by content delivery (due to cache miss) is taken into account. For comparison, we also simulate a broadcasting strategy and a traditional transmission strategy without pre-caching.

We consider $M=19$ small cells each with radius $D = 50$ m hexagonally placed in a macro cell with radius $250$ m. The path-loss model is $30.5+36.7\log_{10}(d)$, where $d$ is the distance between BS and MS in meter \cite{TR36.814}, and $\sigma^2 = -165+10\lg(W_{\max})=-95$ dBm. The
small scale channel is subject to Rayleigh block fading. The bandwidth is $W_{\max}= 10$ MHz, and the maximal transmit power of SBS is $p_{\max}=0.2$ W. The circuit power consumptions in active and sleep modes are $p_{\rm act}=3$ W and
$p_{\rm sle}=1$ W, respectively. The power amplifier efficiency  is $\xi=8\% $ \cite{earth2010}. The duration of each time slot is $\Delta_t = 10$ ms, and each frame contains $T_s=100$ time slots. Each SBS serves RT service arrived with average rate of $\lambda_{\rm RT} = 0.2$ requests per time slot, and the service time follows exponential distribution with average
two time slots. $20\% W_{\max}$ and $20\% p_{\max}$ are reserved for each RT request.

The file size is set as $F = 30$ MBytes \cite{NAA13}. The {file catalog} $\mathcal{N}_f$ that all the users in the macro cell may request contains $N_f=10000$
files, where the files are indexed according to popularity. The user requests follow Zipf distribution with parameter $\beta_f\in [0,1]$, where
$\beta_f$ reflects the ``peakiness" of the common interests of the users.
To model  the uncertainty of the prediction for each user's personal interest, the number of files that each user may request is set as $N_s = 100$, which is a
subset of the {file catalog}  $\mathcal{N}_f$ \cite{LSZ10}, and the subsets for different users may differ. Moreover, the request of each user follows Zipf distribution
with parameter $\beta_s\in [0,1]$.

Each SBS pushes files to 10 MSs in its cell. Each MS moves along a line with random direction and constant speed of $1$ m/s in a small cell,  with minimal distance of $5$ m to $40$ m from the SBS. During the peak time, the content delivery request  arrives at the SBS with average rate $\lambda_{\rm
	CD} = 1.2$ Mbps per MS.
Since simulating the whole day takes long time, we choose $120$ s from the off-peak time to calculate the energy consumed by pushing and
$60$ s from the peak time to calculate the energy consumed by delivering the requested files. The results are obtained
from 1000 Monte-Carlo trails, where the trajectory of each MS stays the same and the small-scale channel and
RT requests vary in every trail.
Unless otherwise specified, this simulation setup is used for all results.

The strategies to be compared are detailed as follows.
\begin{itemize}
	\item Pre-caching with broadcast (with legend ``Broadcast"):  With the predicted common interests for all users in a macro cell, the MBS caches the most popular $N^c_b = 10$ files from $\mathcal{N}_f$ to each user by broadcast once a day during the off-peak time. Since the energy consumed by broadcasting is negligible, we ignore its energy consumption.
	\item Pushing with unicast  (with legend ``Unicast"): With the predicted personal interest of each user, the proposed strategy  pushes the most possibly requested $N^c_u=10$ files from the $N_s$ files to each user by unicast  during the off-peak time. Since the energy consumed by resource planning is  negligible, we ignore the energy consumption at the MBS.
    \item Baseline  (with legend ``Baseline"): The files are transmitted by the SBSs after they are requested by the MSs.
	\end{itemize}

\begin{figure}[!htb]
	\centering
	\includegraphics[width=0.45\textwidth]{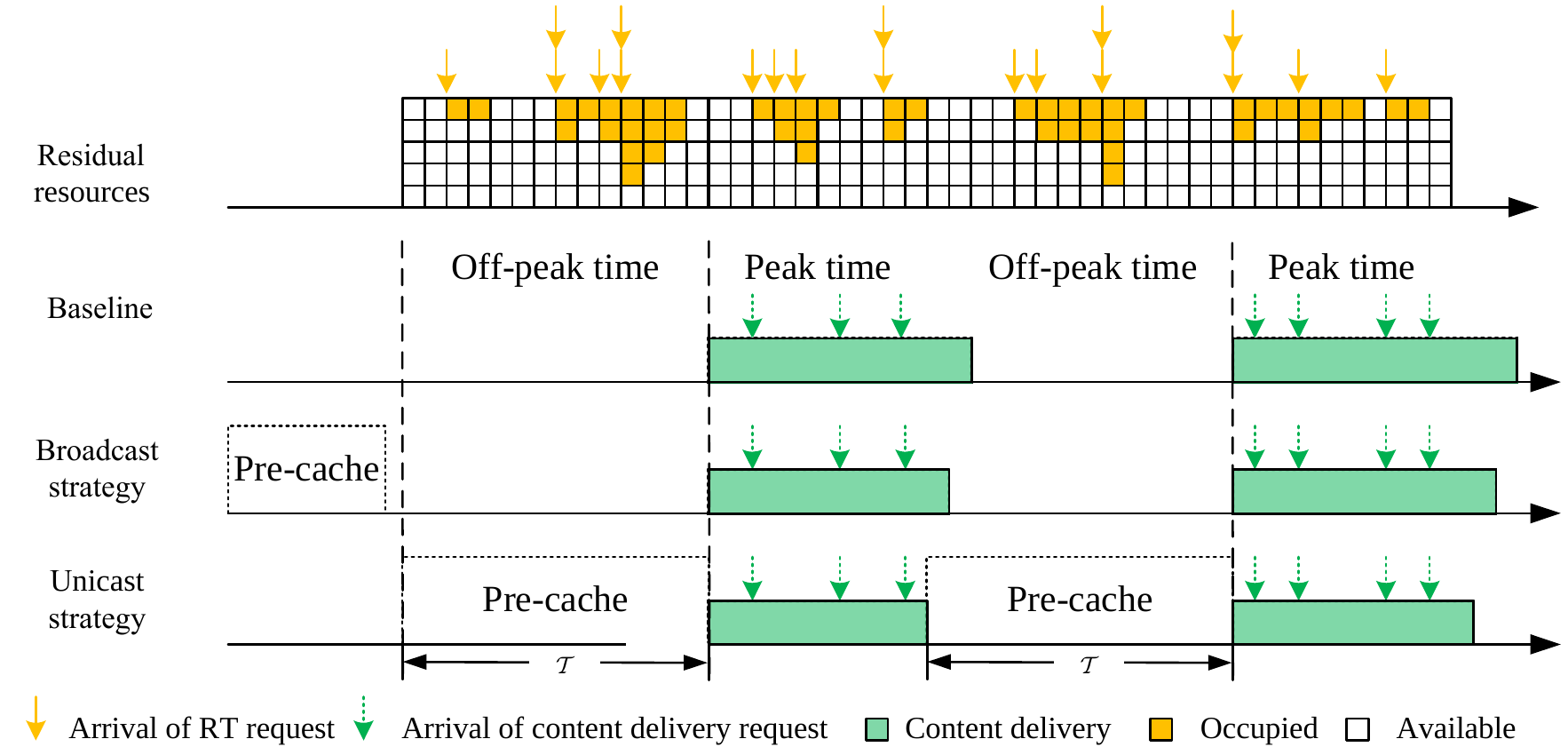}\\
	\caption{Illustration of the simulated strategies. }\label{F:NIPP}
	\vspace{-0.15cm}
\end{figure}

\begin{figure}[!htb]
	\centering
	\includegraphics[width=0.42\textwidth]{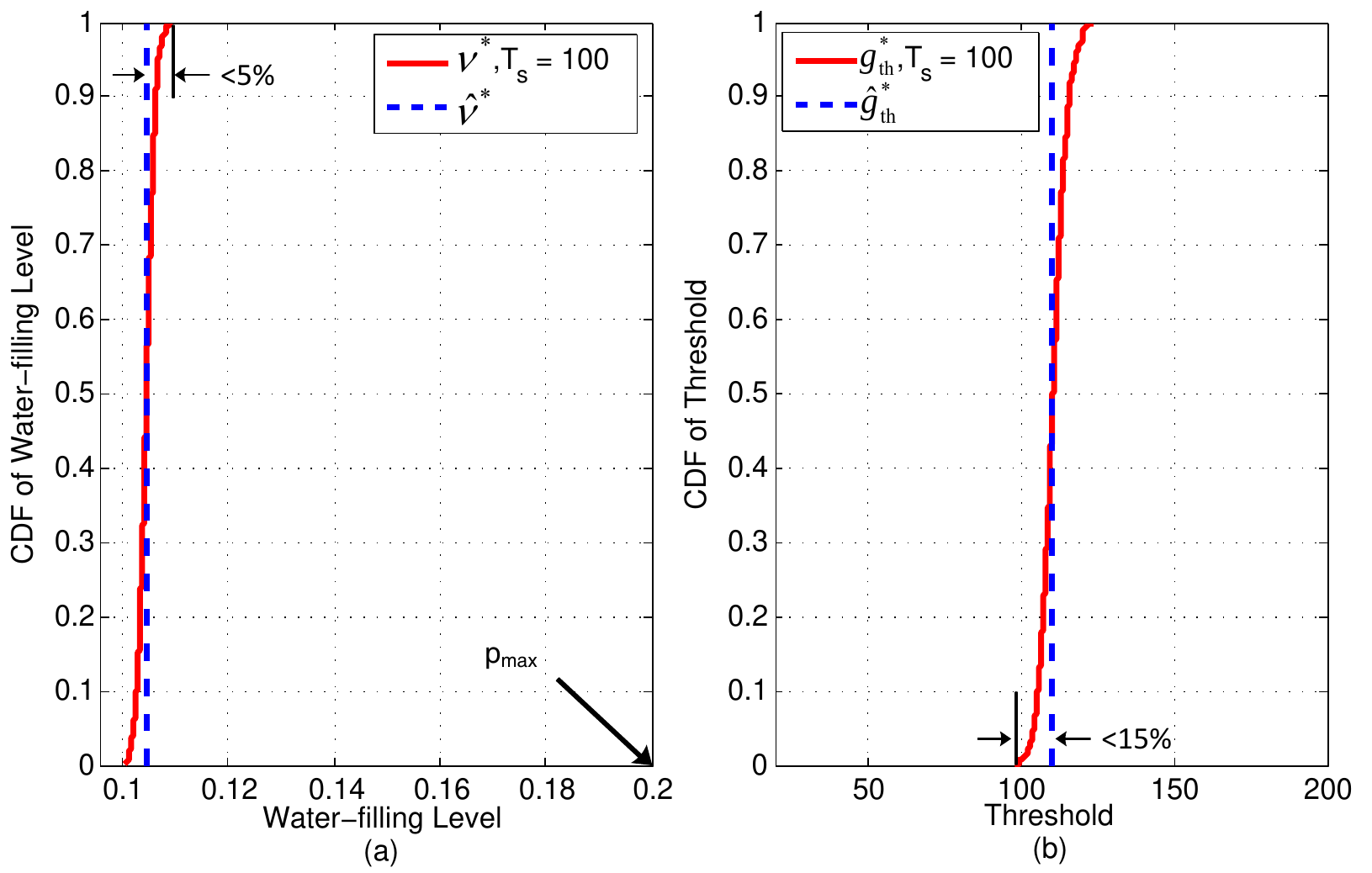}\\
	\caption{ (a) CDF of $\hat \nu^*$ and  $\nu^*$, (b)  CDF of $\hat g^*_{\rm th}$ and  $g^*_{\rm th}$ }\label{F:Estimate}
	\vspace{-0.25cm}
\end{figure}

We first show that the conditions in propositions \ref{Pro:1} and \ref{Pro:2} are easy to satisfy. Fig. \ref{F:Estimate} shows the cumulative distribution function (CDF) of the water-filling level $\nu^*$ and threshold $g^*_{\rm th}$ obtained from problem \eqref{P:1} and $\hat \nu^*, \hat g^*_{\rm th}$ obtained from problem {\bf P1}, where $100$ MBits of content needs to pre-download to a user during $120$ s. We can see that $\nu^*<p_{\max}$ is easy to hold, and
$ |\nu^*-\hat \nu^*|/\hat \nu^*<5\%$ and $|g^*_{\rm th}-\hat g^*_{\rm th}|/\hat g^*_{\rm th}<15\%$ even when $T_s =100$, which is far from infinity.

\begin{figure}[!htb]
	\centering
	\includegraphics[width=0.42\textwidth]{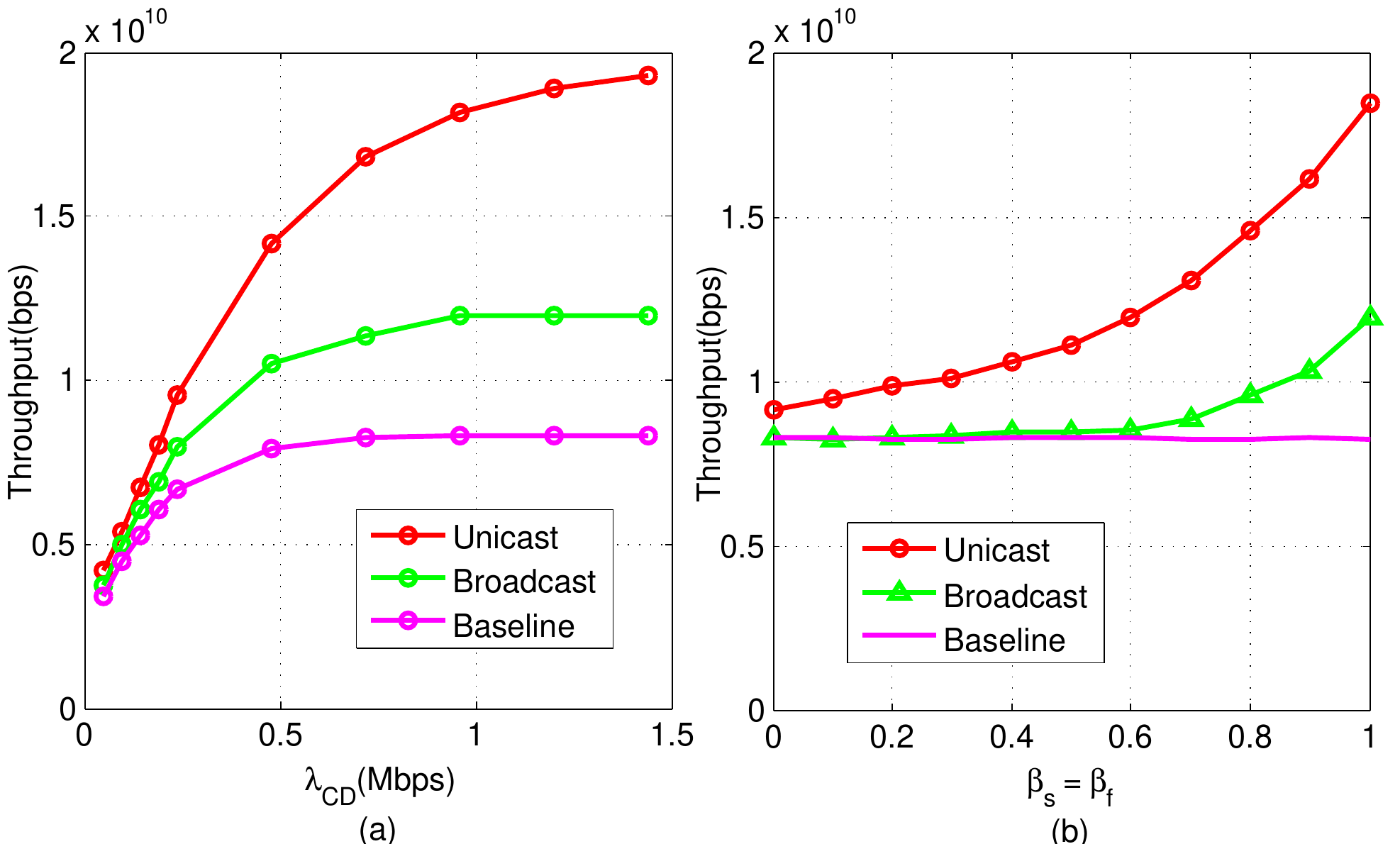}\\
	\caption{Throughput vs. (a) $\lambda_{\rm CD}$, $\beta_f=\beta_s=1$ and (b) $\beta_f=\beta_s$, $\lambda_{\rm CD}$ = 1.2 Mbps}\label{fig.4}
	\vspace{-0.25cm}
\end{figure}

\begin{figure}[!htb]
	\centering
	\includegraphics[width=0.42\textwidth]{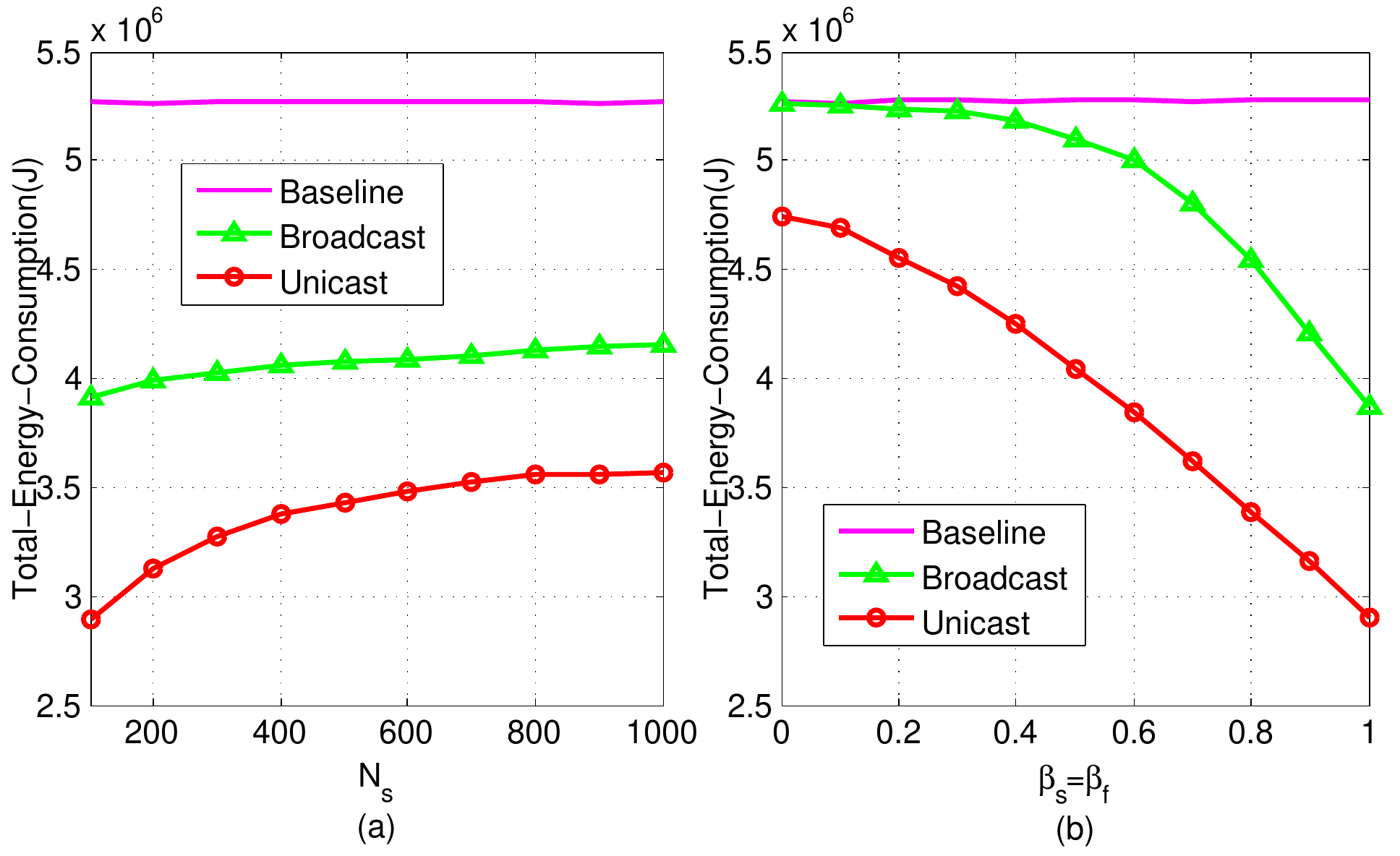}\\
	\caption{Energy consumption vs. (a) $N_s$, $\beta_f=\beta_s=1$, $N_f=100N_s$ and (b) $\beta_f=\beta_s$, $N_s =100$ }\label{fig.5}
	\vspace{-0.25cm}
\end{figure}

In Fig. \ref{fig.4}, we show the network throughput versus the average arrival rate of the content delivery traffic and  $\beta_f=\beta_s$. We can see that the throughput gain of pushing files based on each user's personal interest over pre-caching popular files to all users increases with
$\lambda_{\rm CD}$.
As expected, the throughput gain from ``Broadcast'' and ``Unicast" over the baseline increases with $\beta_f=\beta_s$. When $\beta_f$ is small, ``Broadcast''  achieves the same throughput
as the ``Baseline'', while  ``Unicast"  can improve throughput in all cases.

In Fig. \ref{fig.5}, we show the total energy consumed for content placement and delivery versus $N_s$ and  $\beta_f=\beta_s$. We can observe a surprising result that the proposed unicast strategy consumes less energy than broadcast. This is because more users can fetch their requested files in their own caches with pushing. Yet the energy-saving gain reduces with $N_s$ since a large value of $N_s$ indicates a high prediction uncertainty of each user's interest, which leads to high cache miss rate and hence more energy for delivering the files. More energy can be saved by ``Broadcast'' and ``Unicast" for larger $\beta_f=\beta_s$ with respect to the ``Baseline'', because
the files cached at users have higher probability to be requested. When $\beta_f$
is small, the ``Broadcast''  strategy consumes the same energy as the ``Baseline'', while the proposed ``Unicast" strategy can save energy in all cases.

\section{Conclusion}\vspace{-1mm}
In this paper, we proposed a context-aware unicast pushing strategy based on
the personal interest of each user, aimed at minimizing the extra energy consumed at the BSs for the file pre-downloading. Simulation results showed that the proposed
strategy can improve network throughput and save energy compared to pre-caching with broadcast and traditional network without local caching when the files popularity is
not uniform, even with the demand prediction uncertainty.

\vspace{-2mm}

\bibliographystyle{IEEEbib}
\bibliography{IEEEabrv,2014YCT}
\end{document}